\documentclass[11pt,a4paper]{article}

\usepackage{amsmath}
\usepackage[top=3cm, bottom=3cm, left=2.5cm, right=2.5cm]{geometry}
\usepackage[colorlinks=true, allcolors=blue]{hyperref}
\usepackage{graphicx}
\usepackage{caption}
\usepackage{array}
\usepackage{tabularx}

\usepackage{cite}
\usepackage{lineno}
\usepackage{pdfpages}
\usepackage{setspace}
\usepackage{comment}

\usepackage{xcolor}

\title{Identifying dynamical network markers of \\financial market instability}

\author{Mariko I. Ito$^1$, Hiroyuki Hasada$^1$, Yudai Honma$^{1\ast}$, Takaaki Ohnishi$^2$, \\Tsutomu Watanabe$^3$, Kazuyuki Aihara$^4$}
\date{}

\begin{document}

\begin{spacing}{1.0}
\maketitle

\noindent $^{1}$ Institute of Industrial Science, The University of Tokyo, Tokyo, Japan\\
\noindent $^{2}$ Graduate School of Artificial Intelligence and Science, Rikkyo University, Tokyo, Japan\\
\noindent $^{3}$ Professor Emeritus, Graduate School of Economics, The University of Tokyo, Tokyo, Japan\\
\noindent $^{4}$ International Research Center for Neurointelligence, The University of Tokyo Institutes for Advanced Study, The University of Tokyo, Tokyo, Japan\\
\noindent $^{\ast}$ yudai@iis.u-tokyo.ac.jp

\end{spacing}

\begin{abstract}
Market instability has been extensively studied using mathematical approaches to characterize complex trading dynamics and detect structural change points. 
This study explores the potential for early warning of market instability by applying the Dynamical Network Marker (DNM) theory to order placement and execution data from the Tokyo Stock Exchange. 
DNM theory identifies indicators associated with critical slowing down---a precursor to critical transitions---in high-dimensional systems of many interacting elements. 
In this study, market participants are identified using virtual server IDs from the trading system, and multivariate time series representing their trading activities are constructed. 
This framework treats each participant as an interacting element, thereby enabling the application of DNM theory to the resulting time series. The results suggest that early warning signals of large price movements can be detected on a daily time scale. 
These findings highlight the potential to develop practical DNM-based early-warning systems for large price movements by further refining forecasting horizons and integrating multiple time series capturing different aspects of trading behavior.
\end{abstract}

\maketitle

\section*{Introduction}\label{sec:intro}

Financial instability involves the accumulation of imbalances, such as asset bubbles, resulting in significant fluctuations in asset prices. This boom/bust phenomenon results in severe financial distress, including credit collapse, disruption of financial intermediation, and material deterioration of economic growth and welfare in the real economy~\cite{bisias2012survey,jackson2021systemic}. 
Nowadays, not only academic researchers but also policymakers have been focusing on methodology based on complex systems science to capture the structure of the financial system in which imbalances are embedded elsewhere~\cite{bisias2012survey,battiston2016complexity,jackson2021systemic}.

Complex systems science provides tools for analyzing collective behavior arising from interactions among numerous elements, revealing organizing principles across phenomena such as opinion polarization, epidemic spreading, neuronal synchronization, and climate dynamics~\cite{kwapien2012physical}. Financial markets are recognized as complex systems, and patterns in market behavior, such as price fluctuations and transaction dynamics, have been extensively analyzed. As Battiston et al. (2016) suggested, the complex systems perspective significantly contributes to addressing instability in financial markets~\cite{battiston2016complexity,jackson2021systemic,may2008ecology,saracco2016detecting,abad2016shedding,bardoscia2021physics,yellen2013}.

Analyses of financial markets using complex systems science have uncovered structural patterns in market behavior. For instance, correlations between stock price fluctuations tend to strengthen during financial crises. To capture these correlation structures, the literature proposes methods that represent them as networks, where each node represents a stock and each link denotes a strong correlation between stock price changes~\cite{bonanno2001high, kwapien2012physical, junior2012correlation}. Additionally, a study illustrated transactional relationships among market participants with a network, showing significant heterogeneity and complexity in their interaction structure~\cite{musciotto2021high}.

For complex systems with numerous interacting elements, the Dynamical Network Markers (DNM) theory detects early warning signals of critical transitions---qualitative changes in system behavior, such as the onset of instability. 
In many dynamical systems, critical slowing down occurs prior to a critical transition~\cite{scheffer2009early,drake2010early}. 
DNM theory extends this concept to higher-dimensional dynamics, where interacting elements form a network~\cite{aihara2022dynamical,chen2012detecting}. 
The theory identifies elements associated with critical slowing down that serve as markers for predicting future transitions. 
Thus, DNM theory enables proactive forecasting of critical transitions rather than merely detecting them post-occurrence. 
When applied to biological systems, it is well known as Dynamical Network Biomarkers (DNB)~\cite{aihara2022dynamical,chen2012detecting}. 
For instance, a set of genes can be identified as a DNB within a gene regulatory network. 
Empirical and theoretical studies validate DNB theory's effectiveness in detecting pre-disease states~\cite{chen2012detecting,koizumi2019identifying,aihara2022dynamical}. 
For example, a study demonstrated that a DNB-based indicator significantly increased prior to lung injury from toxic inhalation, successfully identifying the pre-disease state. Additionally, compounds highlighted by the DNB were crucial in the underlying complex pathological processes~\cite{chen2012detecting}. These findings demonstrate that DNM theory effectively forecasts system instability early by identifying elements in key positions within complex interactions.

Given that a market comprises the trading activities of individual participants, applying DNM theory to predict financial instability is possible; however, such analyses require detailed datasets to capture their distinctive trading behavior. 
In financial markets, prices change as a result of interactions in order flow; for example, transactions and the resulting price movements can trigger further orders and transactions~\cite{bikhchandani1998learning,hawkes2018hawkes,wheatley2019endo,bacry2015hawkes,ito2024exogenous,chamley2004rational,ellul2011regulatory}.
Order placements are made by individual participants, but available data typically reflect only aggregated outcomes, such as price changes. Consequently, previous studies on financial markets have focused on the fluctuations of price indices, individual stock prices, and related variables to investigate market conditions~\cite{kwapien2012physical,bacry2015hawkes,bonanno2001high}. 
Access to detailed datasets capturing participants' trading activities, which correspond to the elementary processes underlying price formation, would enable exploration of novel relationships between market instability and participant behavior using DNM theory.

This study examined detailed data to identify trading activities from each virtual server on the Tokyo Stock Exchange (TSE) and assessed the potential to develop early warning signals of market instability based on DNM theory. 
The data included hashed IDs of the virtual servers for each trading activity. 
A market participant was defined as a set of virtual servers exhibiting similar trading behavior. 
We constructed time series representing each participant's trading characteristics and derived the DNM index from these series. 
Our research question focused on predicting significant price fluctuations as manifestations of market instability using the DNM index. 
Thus, this study aimed to forecast instability by observing high-granularity participant-level behavior. 
The analysis indicated that the DNM index can detect early warning signals of significant price fluctuations within days. While further refinement is necessary for practical application, these results represent a crucial step toward early detection of financial market instability using DNM theory, considering the complex interactions among participants' trading activities.

%%%%%%%%%%%%%%%%%%%%%%%%%%%%%%%%%%%%%%%%%%%%%%%%%%%%%%%%%
\section*{DNM theory}\label{subsec:DNM}
First, we outline DNM theory, which proposes a method for detecting early warning signals of system instability~\cite{aihara2022dynamical,chen2012detecting,koizumi2019identifying,liu2015identifying,oya2014forecasting,liu2016personalized}.
In DNM theory, we consider a high-dimensional dynamical system where multiple elements interact: 
\begin{equation}
\mathbf{X}(t)=(X_1(t),~...,~X_N(t)),
\label{eq:time-series}
\end{equation}
where the state of each element evolves as 
\begin{equation}
X_i(t+1) = f_i(\mathbf{X}(t);\beta_i)+\xi_i(t).
\label{eq:dyn_sys}
\end{equation}
Here, the parameter $\beta_i$ represents an environmental factor affecting the state $X_i(t)$. 
The random variable $\xi_i$, following a normal distribution with mean $0$ ($\xi_i\sim N(0,\sigma)$), denotes the noise in the dynamics of state $X_i(t)$. 
DNM theory shows that as the system approaches a critical transition, a set $I$ emerges such that 
(a) the state $X_i(t)$ exhibits large fluctuations for $i\in I$, 
(b) correlations between states $X_i(t)$ and $X_j(t)$ become strong, either positively or negatively, for $i,~j\in I$ ($i\neq j$), and (c) correlations between $X_i(t)$ and $X_j(t)$ weaken when $i\in I$ and $j\notin I$.
These propositions extend the theoretical framework of critical slowing down, originally developed for one-dimensional systems, to high-dimensional systems~\cite{chen2012detecting,aihara2022dynamical}.
We denote the set $I$ the {\it DNM set}.

Empirical studies adopting DNM theory have proposed methods to identify the DNM set from confoundedly interacting elements~\cite{chen2012detecting,aihara2022dynamical,koizumi2019identifying,oya2014forecasting,liu2015identifying,liu2016personalized}. 
DNM theory applies to multivariate time series $\mathbf{X}(t)$ generated by an underlying dynamical system (Eq.~\ref{eq:dyn_sys}) in a data-driven way, without requiring the exact form of the function $f$ for analysis.
In practical DNM analysis, for a subset of elements $U$($\subset \{1,...,N\}$), four indicators are considered:
the average fluctuation of time series, $\mathrm{SD}_U$, over the elements in $U$;
the average correlation strength among elements within the set $U$, $\mathrm{PCC}_U$; the average correlation strength between elements in $U$ and those outside of it, $\mathrm{PCC}_{\overline{U}}$; and the composite indicator $D_U$.
Details of these indicators are provided in the Methods section.
If $U$ corresponds to the true DNM set, then as the system approaches the instability, $\mathrm{SD}_U$ should increase markedly, $\mathrm{PCC}_U$ should approach $1$, $\mathrm{PCC}_{\overline{U}}$ should approach $0$, and $D_U$ should take on a large value indicating collective large fluctuation.
These features disappear once the system surpasses the critical transition, with quantities such as $\mathrm{SD}_U$ and $\mathrm{D}_U$ returning to their normal levels.
Therefore, in empirical studies, the DNM set $I$ is typically identified as the subset $U$ that maximizes $D_U$ prior to instability, based on historical data.
In this paper, we refer to $\mathrm{SD}_U$, $\mathrm{PCC}_U$, $\mathrm{PCC}_{\overline{U}}$ and $D_U$ as the {\it DNM indicators}.
While $D_I$ is commonly used in DNM analysis, $\mathrm{SD}_I$ is frequently emphasized for its robustness in capturing early warning signals~\cite{masuda2024anticipating,nakagawa2016early}.
Once the DNM set $I$ is identified from past data, the corresponding DNM indicator(s) $\mathrm{SD}_I$, $\mathrm{PCC}_I$, $\mathrm{PCC}_{\overline{I}}$ and $D_I$ derived from $I$ can be used to predict future instabilities.

In this study, we examine the potential of DNM theory for predicting future instability in financial markets by applying it to multivariate time series data constructed from the trading activities of participants. Each participant ($i$) is regarded as an element of the system, and we construct a time series $x_i(t)$ to quantify its trading behavior. 
Given the influence of order placements or executions on one another, or herding behavior, in financial markets~\cite{bikhchandani1998learning,chamley2004rational,ellul2011regulatory}, we assume that the temporal changes in a participant's trading behavior, $x_i(t)$, may be influenced by the behavior of another participant, $x_j(t)$. 
Thus, we consider a system where each element's state, i.e., the trading behavior of each participant, evolves over time through mutual interaction, implying a function $f$ that represents these interactions, forming a dynamical system as in Eq.~(\ref{eq:dyn_sys}). Based on this framework, we apply DNM theory to the multivariate time series $\mathbf{x}(t)=(x_1(t),...,x_N(t))$ and evaluate its ability to predict market instability.

%%%%%%%%%%%%%%%%%%%%%%%%%%%%%%%%%%%%%%%%%%%%%%%%%%%%%%%%%
\section*{Application of DNM theory to financial time series}\label{sec:Application of DNM analysis to financial time series}

This study analyzes detailed data on order placements and executions in the Tokyo Stock Exchange from 5 Nov 2019 to 31 Dec 2020, encompassing the global financial instability triggered by the COVID-19 pandemic especially in February and March 2020.

We define a participant, construct time series $x_i(t)$ representing the trading activity of each participant, and apply DNM theory to these time series, as follows.

\subsection*{Market participant}
We define a market participant based on the virtual server IDs (VSIDs) in the dataset. 
A VSID is an identification number assigned to each TSE customer who submits orders through its trading system~\cite{goshima2019trader,ohyama2022_jafee_e}. 
The identity of the member associated with each VSID is anonymized. 
Multiple VSIDs can be involved in a single sequence of trading activity, as shown when different VSIDs submit new, change, and cancellation orders that share the same order ID~\cite{goshima2019trader,sato2023inferring,yamada2023_hft_e}. 
We define a participant as a set of VSIDs, considering the concept of a {\it trading desk}, order ID sharing, and similarities in trading characteristics among VSIDs, as detailed in the Methods section. 
We identified a total of $186$ market participants: $68$ HFT (High-frequency trading) type, $25$ broker type, $3$ general investor type, and $90$ other type participants~\cite{yamada2023_hft_e}. Note that our analysis does not aim to recover the exact set of VSIDs associated with a single agent but rather groups VSIDs by similarities in trading patterns to define participants.

\subsection*{Time series representing trading behavior}
In our analysis, we construct a time series $x_i(t)$ for each participant $i$ in each trading session (morning or afternoon) and day.
Each session lasts $2.5$ hours, with a time unit of one minute, yielding a time series of length $T:=150$ (minutes).

\begin{figure}[h]%[tb]
\centering
\includegraphics[width=0.85\textwidth]{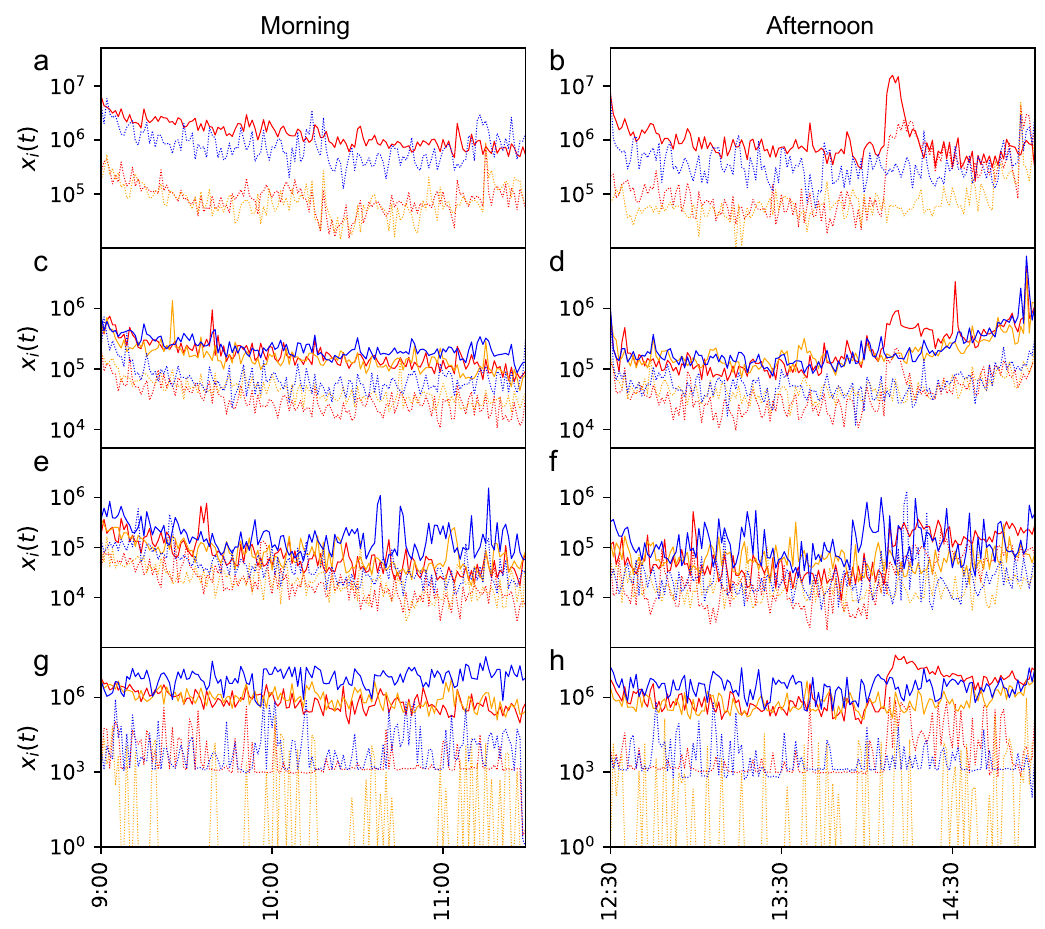}%[width=16cm]{fig1}%Fig/fig_xt}%
\caption{{\bf Examples of time-series evolution.} Panels show the time series $x_i(t)$ of type vol3 (V\_O\_LO\_b) for four participant types: HFT (a,b), broker (c,d), general investor (e,f), and others (g,h). Panels (a,c,e,g) and (b,d,f,h) correspond to the morning and afternoon sessions, respectively. Each panel displays $x_i(t)$ on three representative days: 28 August 2020 (highest Relative Volatility, red), 8 July 2020 (medium Relative Volatility, orange), and 26 February 2020 (lowest Relative Volatility, blue). For each participant type, the time series for the participant with the largest number of orders (solid line) and that with a medium number of orders (dashed line) during the analysed period are shown, provided the participant was active on the focal day.}
\label{fig:xit}
\end{figure}

We categorize the time series into three groups to represent each participant's trading behavior. For each category, we analyse nine distinct time series defined by order placement or execution type, as summarized in Tables~\ref{tab:time_series_vol}--\ref{tab:time_series_pp} in the Methods section: 
(1) time series capturing trading volumes per minute (vol1--9, Table~\ref{tab:time_series_vol}, an example for vol3 is shown in Fig.~\ref{fig:xit}); 
(2) time series representing participant's centrality within the co-trading network, which reflects overlaps in traded securities (co1--9, Table~\ref{tab:time_series_co});
(3) time series describing the trading point process, namely the sequence of times for individual trading activities (pp1--9, Table~\ref{tab:time_series_pp})~\cite{karsai2018bursty,bacry2015hawkes}.

\subsection*{DNM analysis}
By employing the DNM theory, we analysed each time series $x_i(t)$ introduced in the previous section.
We focused on $\mathrm{SD}_I$ among the four DNM indicators, and compared its daily values with large price fluctuation, allowing for time lags of days to weeks.
We chose $\mathrm{SD}_I$ because it is robust against noise~\cite{masuda2024anticipating,nakagawa2016early,chen2012detecting}, while the other indices, $\mathrm{PCC}_I$, $\mathrm{PCC}_{\overline{I}}$ and $D_I$, are affected by strong noise in the cross-correlation between time series in our pre-analysis.
Since $\mathrm{SD}_I$ is the mean standard deviation of time series across participants in the DNM set, we calculated the standard deviation of all participants' time series and selected those whose standard deviation increased prior to large price movements to determine the elements of the DNM set.

For robust comparisons between the standard deviation of the time series and large price movements, we smoothed the daily standard deviation and computed the logarithm of its ratio to that of the previous week, yielding the Log-Relative Standard Deviation (Log-RSD), $R_i(d)$, as detailed in the Methods section.
We later examine whether, restricted to the DNM set, the mean Log-RSD of participants serves as an early-warning signal for large price movements. 

\begin{figure}[h]%[tb]
\centering
\includegraphics[width=0.85\textwidth]{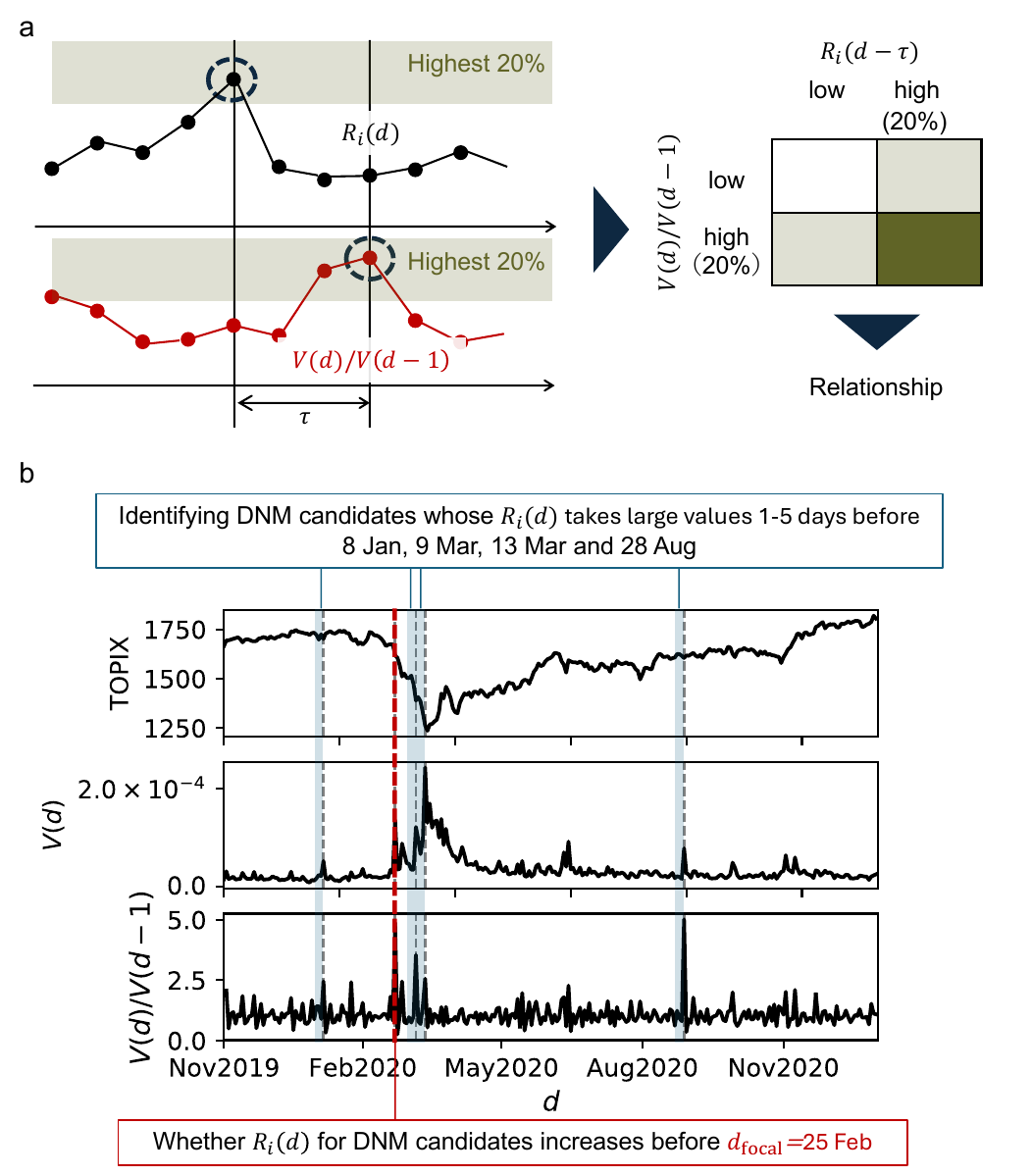}%[width=16cm]{fig2}%Fig/fig_schematic}
\caption{{\bf Procedure of analyses.} (a) Comparison between $R_i(d-\tau)$ and $V(d)/V(d-1)$. (b) Cross-validation for the response of $R_i(d)$ prior to each focal turmoil day.}
\label{fig:schematic}
\end{figure}

In this study, the daily price fluctuation is measured based on the Tokyo Stock Price Index (TOPIX).
We defined the volatility, $V(d)$, on day $d$ as the standard deviation of the logarithmic returns of TOPIX values, focusing on $V(d)/V(d-1)$, termed Relative Volatility, which captures the onset of instability rather than its persistence.
Figure~\ref{fig:schematic}(b) shows the evolution of $V(d)$ and the ratio $V(d)/V(d-1)$.
To examine the performance of the early-warning signal in detecting significant price movement episodes, we identified the five days with the highest Relative Volatility during the observation period.
These {\it turmoil days} are 8 January, 25 February, 9 March, 13 March, and 28 August; consecutive days were excluded from the selection.

%%%%%%%%%%%%%%%%%%%%%%%%%%%%%%%%%%%%%%%%%%%%%%%%%%%%%%%%%
\section*{Results}\label{sec:results}

\subsection*{Relationship between Log-RSD and Relative Volatility for various time lags}
We first compare Log-RSD, $R_i(d)$, with Relative Volatility, $V(d)/V(d-1)$, throughout the observation period.
For various time lags $\tau >0$, we evaluate the relationship between $R_i(d-\tau)$ and $V(d)/V(d-1)$ to examine whether Log-RSD measured $\tau$ days earlier predicts subsequent large price movements.
We asses this relationship using Fisher's exact test, applied to a $2\times 2$ contingency table counting days $d$ satisfying four conditions: ``both $V(d)/V(d-1)$ and $R_i(d-\tau)$ are high''; ``$V(d)/V(d-1)$ is high while $R_i(d-\tau)$ is not''; ``$V(d)/V(d-1)$ is not high while $R_i(d-\tau)$ is high''; and ``neither measure is high''.
A value is considered ``high'' if it falls within the top $20$th percentile of all analyzed days.
For each time-series type, participant $i$ and time-lag $\tau$, we compute the $p$-value of Fisher's exact test to evaluate the null hypothesis---that there is no relationship between $V(d)/V(d-1)$ and $R_i(d-\tau)$.

\begin{figure}[h]%[tb]
\centering
\includegraphics[width=0.85\textwidth]{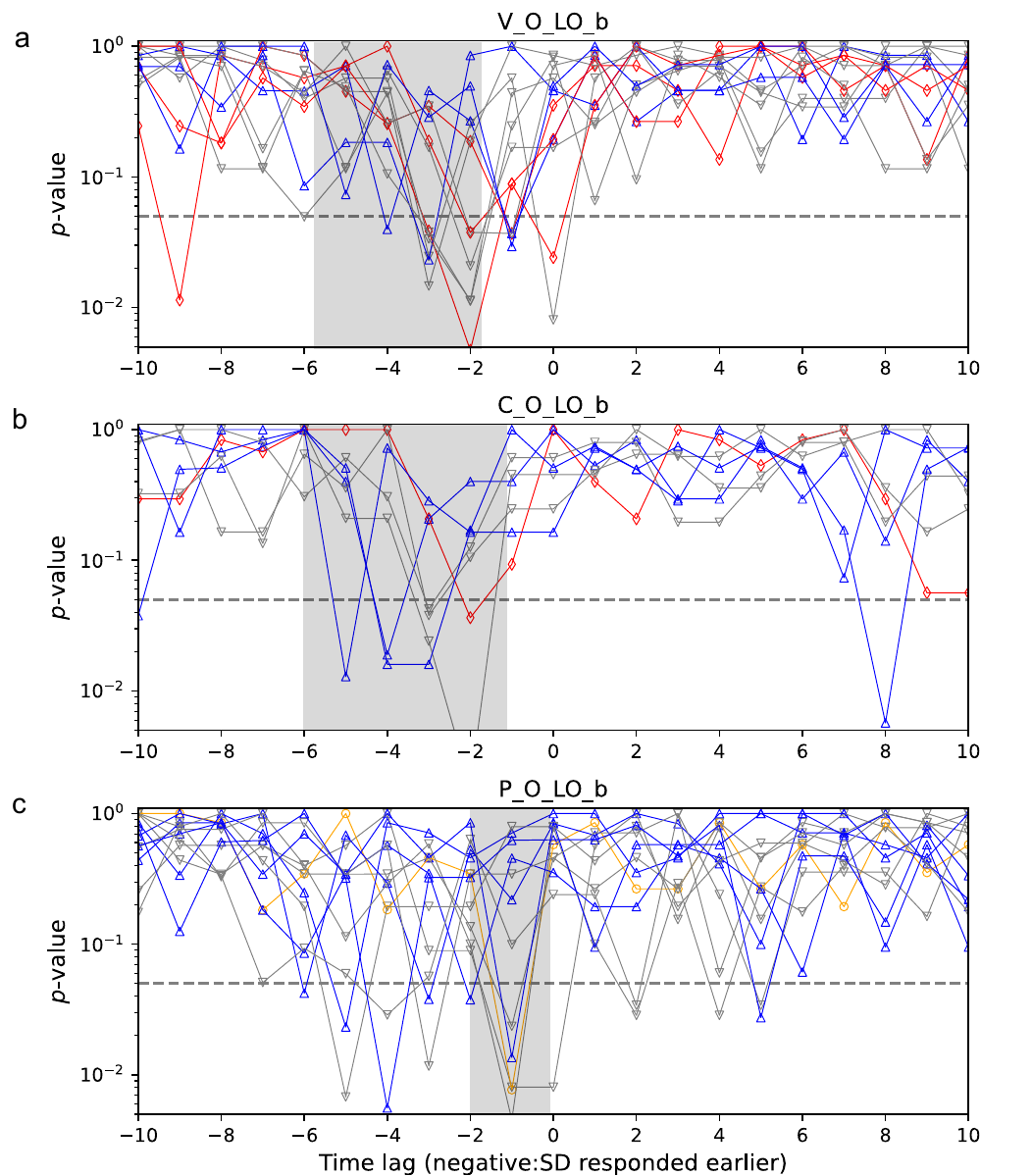}%[width=16cm]{fig3}%Fig/fig_fisher_p.pdf}
\caption{{\bf Relationship between $R_i(d-\tau)$ and $V(d)/V(d-1)$.} 
The vertical axis shows the $p$-value of Fisher's exact test for this relationship, and the horizontal axis shows the time lag $-\tau$. Panels (a), (b), and (c) show the results for the time-series types vol3 (V\_O\_LO\_b), co3 (C\_O\_LO\_b), and pp3 (P\_O\_LO\_b), respectively. Each plot corresponds to a participant who could plausibly belong to the DNM set. A participant was identified as a candidate if its $p$-value fell below $0.05$ (the dashed horizontal line) at least once within $-5\leq -\tau < 0$, and if the mean $p$-value in this interval was lower than that in $0\leq -\tau < 5$. Plots for HFT, broker, general investor, and other participant types are colour-coded in blue, red, orange, and grey, respectively.}
\label{fig:pval}
\end{figure}

We observed a continuous decline in the $p$-value of Fisher's exact test for some participants, indicating a statistically significant association between Relative Volatility and Log-RSD over time lags of one to five days.
Figure~\ref{fig:pval} presents the $p$-value as a function of time lag for selected time-series types, vol3, co3, and pp3 (results for all time-series types are in Section S.3 of Supplementary Information).
The figure includes only participants likely to belong to the DNM set.
A candidate was identified if its $p$-value was below $0.05$ at least once within $-5\leq -\tau < 0$, and if the mean $p$-value in this interval was lower than in $0\leq -\tau < 5$. 
In other words, these participants exhibited Log-RSD patterns associated with subsequent changes in Relative Volatility one to five days in advance, with the association fading after the volatility shift occurred.
In Fig.~\ref{fig:pval} and additional figures in Section S.3 of Supplementary Information, many candidates show fluctuating $p$-values, complicating the discernment of a clear pattern between Log-RSD and Relative Volatility.
In some cases, the $p$-value significantly decreases within the earlier range of $-10\leq -\tau < -5$, possibly reflecting the response of Log-RSD to previous pronounced volatility shifts.
These observations highlight the challenge of consistently assessing the relationship throughout the observation period.
Nevertheless, we identified a subset of candidates whose $p$-values changed smoothly with the time lag and declined markedly within $-5\leq -\tau < 0$.
This result suggests that the Log-RSD of these candidates tends to respond one to five days before the rise in Relative Volatility.

Notably, the time lag at which the $p$-value reached its minimum varied across time-series types and participants, indicating differing lead-lag relationship between Log-RSD and Relative Volatility among participants, with lead times spanning several days.

\subsection*{Relationship between Log-RSD and Relative Volatility}
Subsequently, we examine the validity of the early-warning signal constructed from the DNM set, comprising participants whose Log-RSD, $R_i(d)$, increased several days prior to large price movements.
We employed a cross-validation approach (Fig.~\ref{fig:schematic} (b)).
Each turmoil day served as a focal forecast day, $d_{\mathrm{focal}}$.
For each of the remaining four turmoil days ($d\neq d_{\mathrm{focal}}$), we computed the five-day average Log-RSD, $\overline{R_i(d)}$, over day $d$ and the preceding four days: $\overline{R_i(d)} = \left( R_i(d) + R_i(d-1) + R_i(d-2) + R_i(d-3) + R_i(d-4)\right)/5$.
We then computed the mean of these averages across the four non-focal turmoil days, 
\begin{equation}
\displaystyle\frac{1}{4}\sum_{d\neq d_{\mathrm{focal}}}\overline{R_i(d)}.
\label{eq:evaluation}
\end{equation}
Participants with Eq.~(\ref{eq:evaluation}) values in the top ten and exceeding $1$ were provisionally designated as members of the DNM set, $I_{d_{\mathrm{focal}}}$.
We then evaluated the validity of $I_{d_{\mathrm{focal}}}$ by examining whether $R_i(d)$ of its members increased prior to $d_{\mathrm{focal}}$.
In other words, the provisional DNM set $I_{d_{\mathrm{focal}}}$ was identified based on Log-RSD behavior preceding the four non-focal turmoil days, and its forecasting validity was assessed by testing whether the same behavior occurred before $d_{\mathrm{focal}}$.
Note that $I_{d_{\mathrm{focal}}}$ was determined separately for each time-series type and each $d_{\mathrm{focal}}$.

\begin{figure}[h]%[tb]
\centering
\includegraphics[width=0.85\textwidth]{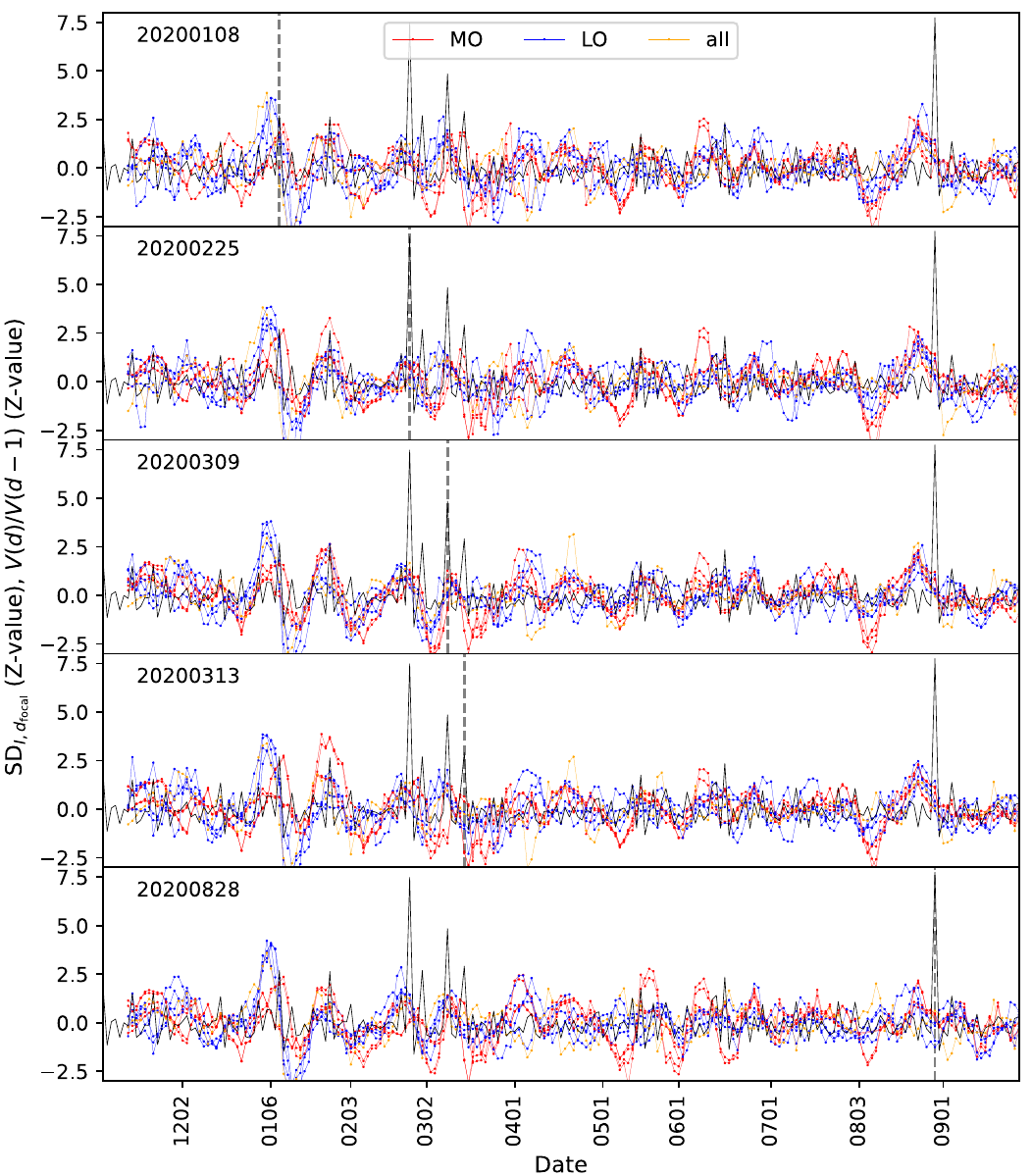}%[width=16cm]{fig4}%Fig/fig_rsd_volume_nsa.pdf}
\caption{{\bf Evolution of $\mathrm{SD}_{I,d_{\mathrm{focal}}}$ derived from time series representing trading volume (vol1--vol9).} Each panel corresponds to a different $d_{\mathrm{focal}}$, indicated by the dashed vertical line. The Z-values of $\mathrm{SD}_{I,d_{\mathrm{focal}}}$ for time-series types considering new orders, executions, and all orders are shown in red, blue, and orange, respectively. The Z-values of $V(d)/V(d-1)$ are shown in black.}
\label{fig:sdidfocal1}
\end{figure}

To examine the validity of the DNM set, $I_{d_{\mathrm{focal}}}$, we constructed a DNM indicator $\mathrm{SD}_{I,d_{\mathrm{focal}}}$ by averaging $R_i(d)$ over participants in $I_{d_{\mathrm{focal}}}$, and evaluated its behavior prior to the turmoil days.
Figure~\ref{fig:sdidfocal1} presents the daily values of $\mathrm{SD}_{I,d_{\mathrm{focal}}}$ for each time-series type in vol1--9 (results for other time series categories are presented in Section S.4 of Supplementary Information).
Each time series is colour-coded: market orders in red, limit orders in blue, and all order types in orange.
Figures in Section S.4 use alternative colour schemes distinguishing trading types (new order or execution) and trade direction (buy or sell) across all time-series types.
Results for several time-series types associated with market orders and co-trading relationships are not shown due to insufficient sample size.

The DNM indicator, $\mathrm{SD}_{I,d_{\mathrm{focal}}}$, tended to rise prior to each focal turmoil day, $d_{\mathrm{focal}}$, across various time-series types, and this tendency appeared largely independent of the specific series considered.
Throughout Fig.~\ref{fig:sdidfocal1} and figures in Section S.4, $\mathrm{SD}_{I,d_{\mathrm{focal}}}$ generally peaked before $d_{\mathrm{focal}}$, particularly for $d_{\mathrm{focal}}=\text{8 Jan}$ and $\text{28 Aug}$.
In February, when the COVID-19 pandemic began affecting the market, $\mathrm{SD}_{I,d_{\mathrm{focal}}}$ rose sharply from a relatively low baseline, although its peak level remained modest in absolute terms.
For $d_{\mathrm{focal}}=\text{9 Mar}$ and $\text{13 Mar}$, amid COVID-19 shock, $\mathrm{SD}_{I,d_{\mathrm{focal}}}$ fluctuated widely, making it difficult to identify a clear pre-instability rise, presumably due to these two turmoil days occurring in close succession and the slower response time scale of $\mathrm{SD}_{I,d_{\mathrm{focal}}}$ compared to rapid fluctuations in Relative Volatility during this period.
$\mathrm{SD}_{I,d_{\mathrm{focal}}}$ associated with limit orders tended to increase earlier than that for market orders on $d_{\mathrm{focal}}=\text{8 Jan}$.
However, such timing differences were not consistently observed across all time-series types.
Overall, the rise of $\mathrm{SD}_{I,d_{\mathrm{focal}}}$ prior to several $d_{\mathrm{focal}}$ events across diverse time-series types supports the effectiveness of the DNM indicator $\mathrm{SD}_{I,d_{\mathrm{focal}}}$ as an early-warning signal, although systematic comparison among time-series types---or potential aggregation across them---remains an important direction for future work.

Our DNM indicator captures behavioral shifts in trading driven by institutional or structural market factors.
For example, the focal turmoil day $d_{\mathrm{focal}}=\text{8 Jan}$ occurred just before the Special Quotation Day (SQ Day) on 10 January~\cite{JPX_SQ_Glossary}.
Accordingly, the DNM indicator likely reflected changes in participants' trading behavior in anticipation of the upcoming SQ Day.
This illustrates that our DNM analysis, developed purely from data, can detect early signals of trading adjustments without explicitly incorporating institutional or external information such as scheduled events or news.

\subsection*{Characteristics of market participants comprising the DNM set}
\begin{figure}[h]%[tb]
\centering
\includegraphics[width=0.85\textwidth]{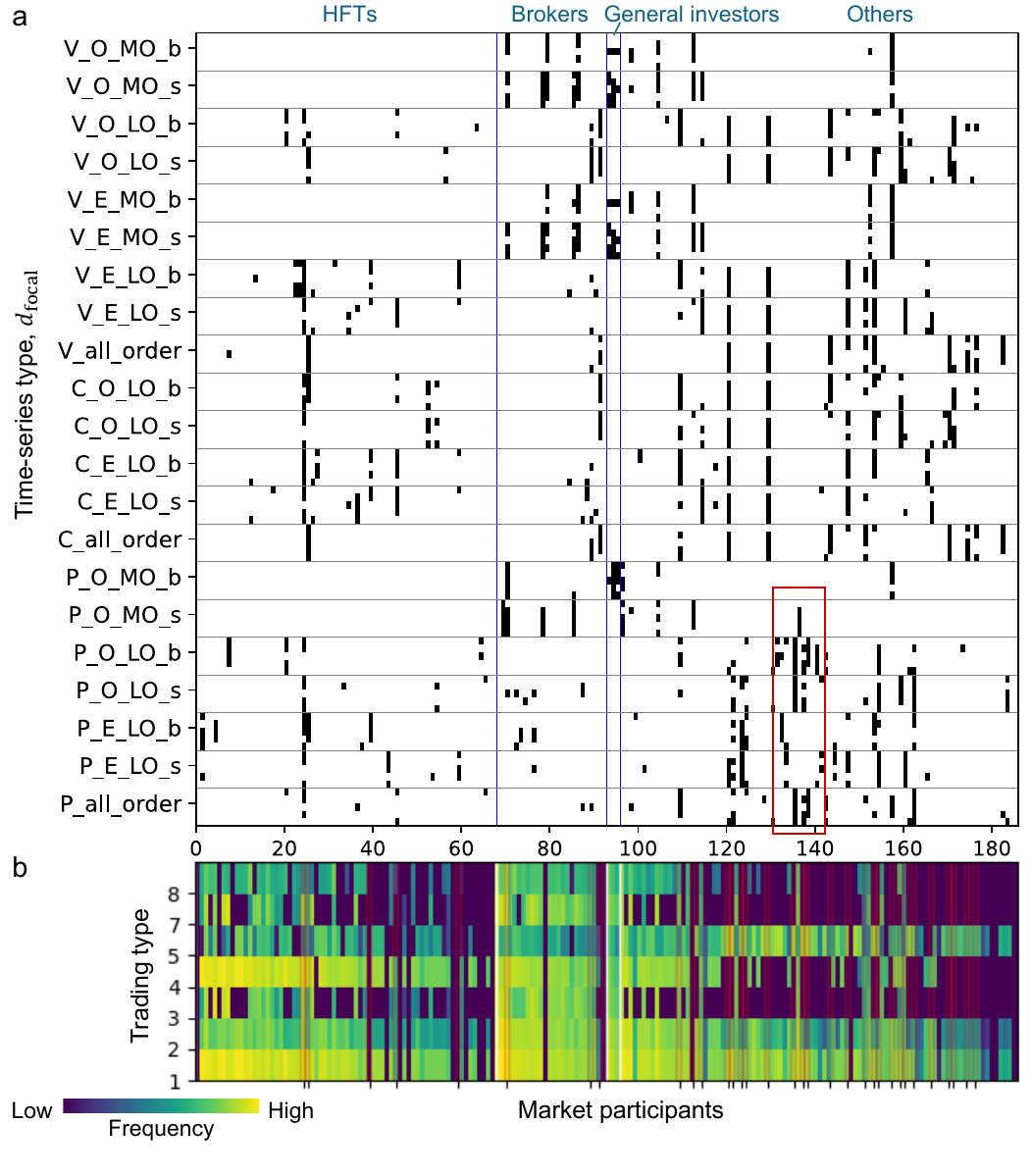}%[width=15cm]{fig5}%Fig/fig_dnm_elements_rev.pdf}
\caption{{\bf Participants comprising $I_{d_{\mathrm{focal}}}$.} In both panels (a) and (b), the horizontal axis lists participants ordered first by types and then in descending order of their total number of order placements within each type. Participants $1$--$68$, $69$--$93$, $94$--$96$, and $97$--$186$ correspond to HFT-type, Broker-type, General-investor-type, and Others-type participants, respectively. (a) The vertical axis lists all focal turmoil days $d_{\mathrm{focal}}$ for each time-series type. A filled cell indicates that the corresponding participant belongs to $I_{d_{\mathrm{focal}}}$ for that time-series type and focal day. The horizontal gray lines separate different time-series types. The red rectangle highlights participants that appear in $I_{d_{\mathrm{focal}}}$ specifically for time-series types representing trading point processes. (b) The vertical axis lists trading types as follows: 1: New order, 2: Execution, 3: Change, 4: Cancellation, 5: Expiration, 7: Cancellation and new order, 8: Closing transactions. The heatmap colour represents the frequency of order placements or executions in January 2020 for each trading type and participant, normalized by the mean value of each row. Principal participants are highlighted with red frames; their positions are indicated by tick marks on the horizontal axis, and white vertical lines separate different participant types. }
\label{fig:marketparticipants_all}
\end{figure}

We observed substantial overlap among participants in the DNM set, $I_{d_{\mathrm{focal}}}$, across various focal turmoil days.
Figure~\ref{fig:marketparticipants_all} (a) summarizes the composition of $I_{d_{\mathrm{focal}}}$ by focal turmoil day $d_{\mathrm{focal}}$ and time-series types (detailed summaries are in Section S.5 of the Supplementary Information). Market participants are arranged in descending order of total order placements and categorized by types (HFT, broker, general investor, and others) along the horizontal axis, with focal turmoil days displayed side by side for each time-series type along the vertical axis. A filled cell indicates that the participant belongs to $I_{d_{\mathrm{focal}}}$ for the given combination of time-series type and focal turmoil day.
The plot reveals many vertical segments, indicating that certain participants consistently belong to the DNM set across multiple turmoil days.
This suggests the robustness of the DNM set and its potential to capture early-warning signals for future market instabilities based on participants' trading behavior preceding past instability events.

We compared the DNM set members across different time-series types and found that some market participants were repeatedly identified across multiple types, while others were specific to particular ones.
As shown in Fig.~\ref{fig:marketparticipants_all} (a), several participants belonged to DNM sets across various time series types, including buy and sell orders, trading volumes (vol1--9), co-trading relationships (co1--9), and point processes (pp1--9). Notably, some participants, particularly those classified as ``other''-type participants, were found specifically in the DNM sets associated with time series capturing trading point processes, as highlighted by the red rectangle.
These participants are identified as DNM members primarily when trading behavior is represented by the precise timing of individual trading activities rather than aggregated activity per minute.
Furthermore, whether a participant belongs to a DNM set appears largely independent of their total number of order placements, as indicated by filled cells distributed across a wide range of horizontal positions.

We examined the characteristics of {\it principal participants}---those appearing repeatedly in the DNM set across multiple time-series types and focal turmoil days---and found that a broker-type principal participant consistently occupied a central position in the co-trading network.
Participants belonging to several DNM sets can be regarded as key agents in the DNM framework.
We counted the number of combinations of time-series types and $d_{\mathrm{focal}}$ in which each participant appeared in the corresponding DNM set, identifying the top ten by this count as principal participants; another ten were selected based on point process types  (pp1--9).
Figure~\ref{fig:marketparticipants_all} (b) presents the frequencies of order placements and executions for each participant in January 2020, with principal participants highlighted (results for other months are in Section S.6.1 in Supplementary Information).
This figure indicates that principal participants do not display distinctive characteristics in their order placement or execution frequencies.
We analysed their positions in the co-trading network for January 2020 (Fig.~\ref{fig:stock_sharing_network}; results for other months are in Section S.6.2 in Supplementary Information), where nodes represent participants and links indicate strong co-trading relationships. 
The detailed construction of the co-trading network is shown in the Methods section.
Co-trading arises from behavioral interdependence among participants, reflecting these underlying interactions.
Such behavioral interactions underpin the DNM framework; trading behaviors influence one another, and the observed time series are interpreted as outputs of these mutual interactions.
Therefore, analyzing the co-trading network is crucial for understanding the structural basis of DNM-based signals.
As shown in Fig.~\ref{fig:stock_sharing_network}, a broker-type principal participant occupies a central position with a markedly high degree in the backbone co-trading network.

%%%%%%%%%%%%%%%%%%%%%%%%%%%%%%%%%%%%%%%%%%%%%%%%%%%%%%%%%
\section*{Discussion}\label{sec:discussion}
Our analysis applied the DNM theory to complex trading interactions among participants, demonstrating the potential for constructing early-warning signals based on this theoretical framework. 
The DNM theory extracts elements associated with early indicators of system instability from systems composed of interacting components. 
We constructed multivariate time series representing participants' trading activities and applied the DNM theory by treating each participant as an element of the system. 
Specifically, we compared the performance of early-warning signals derived from various time series. 
The results indicate that early signals of large price movements can be detected on a timescale of several days. 
These findings were obtained through a data-driven approach leveraging extensive order placement and execution data based on the DNM theory. 
Our analysis provides a novel perspective for detecting shifts in the macroscopic state of financial markets and suggests that this approach can enhance market stability assessment and risk management.

A key strength of our market instability forecast is its ability to capture early-warning signals of structural changes in trading activities, rather than merely detecting those changes after they occur. 
Previous studies predicting financial crises have focused on identifying structural changes in interactions among financial elements---such as correlations among stock prices or networks of interbank contracts---and forecasting crises that occur after these changes~\cite{zheng2012changes,saracco2016detecting,battiston2016complexity,jackson2021systemic}. 
In contrast, our approach aims for early detection of these structural transitions, enabling anticipation of instability before macroscopic market state shifts. If implemented, such an early-warning system could alert much earlier, providing valuable lead time for risk management.

Realizing a practical forecast of market instability based on the DNM theory requires further examination of how to account for exogenous factors influencing markets. 
Practical forecasting typically demands precision regarding the timing of future instabilities; however, our analyses showed gradual responses, with $\mathrm{SD}_{I,d_{\mathrm{focal}}}$ increasing several days before large price movements. 
While these responses indicate the potential effectiveness of DNM-based forecasting, further work is needed to determine the time lag between the indicator's rise and instability onset, the critical threshold for issuing alerts, and other calibration parameters.
Most prior studies focused on qualitative insights, while those that pursued quantitative forecasting encountered difficulties in achieving precise results~\cite{song2024early}. 
This may stem from noise introduced by exogenous factors influencing trading activities. 
Real financial markets are open systems subject to continual external shocks, deviating from the DNM assumption of closed interactions among participants with limited random noise. 
These exogenous factors significantly affect the DNM indicator and are unavoidable in practice. Thus, developing a DNM-based early-warning indicator that remains robust against such external influences is also an important direction for future study.

Examination of different time scales for DNM analysis is essential to mitigate the influence of exogenous factors.
In our study, we calculated the DNM indicator daily and smoothed it using a moving average with a one-week window, enhancing its robustness against external shocks.
Analyses on shorter time scales---of several minutes---may also be less affected by external factors.
For instance, in the TSE, after the domestic session closes, overseas markets continue trading, with the resulting information incorporated into the TSE at the next morning's opening as external news. 
By adopting a shorter time scale, exogenous influences from overnight developments in foreign markets can be largely eliminated, allowing the analysis to remain self-contained within a single trading day. Additionally, given the prevalence of high-frequency trading, shorter time-scale analyses should be prioritized in future DNM studies of financial markets. 
Furthermore, considering differences in trading speeds among participant types, applying DNM analysis to specific categories, such as HFT types, could provide deeper insights into early-warning phenomena in real markets.

Another future direction is constructing a DNM indicator that aggregates information from multiple time-series types. 
In our study, although there was no significant difference among the time-series types in how strongly $\mathrm{SD}_{I,d_{\mathrm{focal}}}$ responded to market instabilities, the timing of these increases varied slightly across types. 
This finding suggests that each time-series type captures distinct aspects of market instability. 
Aggregating diverse information sources enhances predictive performance, a principle known as collective intelligence, exemplified by ensemble learning methods~\cite{zhou2025ensemble,ito2023casting}. 
Thus, more robust DNM-based early-warning signals could be developed by aggregating different time-series types appropriately.

Lastly, we discuss the characteristics of principal participants and the potential controllability of market instability.
The DNM set comprises elements associated with critical slowing down, a phenomenon preceding instability in complex systems of interacting components. 
A principal participant should be viewed not as an outlier in trading behavior, but as an agent occupying a structurally significant position within the trading interaction network. 
Certain participants may suppress the propagation of shocks caused by external factors, while others influence collective trading dynamics at a systemic level~\cite{jackson2021systemic,Klincic2023PRE,Liu2011Nature}.
The DNM set may thus consist of these structurally important participants within the trading structure.
To examine this idea, we constructed a co-trading network among participants to assess the positions of principal participants within the trading structure. 
While the precise topology of trading interactions cannot be fully captured, co-trading relationships can partially reflect these connections. 
The analysis revealed that a principal participant---specifically, a broker-type participant---occupied a central position in the co-trading network, possessing a notably high degree within its backbone structure. 
This finding suggests that such participants may play key roles in the collective dynamics of market stability.
Future investigations into the positions and roles of principal participants within the trading network could provide insights into controllability of market instability. 
This work would benefit from a macroscopic and integrative perspective, combining various aspects of trading behavior and interaction structure.

%%%%%%%%%%%%%%%%%%%%%%%%%%%%%%%%%%%%%%%%%%%%%%%%%%%%%%%%%
\section*{Conclusion}\label{sec:conclusion}
In this study, we examined the construction of early-warning signals for market instability by applying the DNM theory to trading data from the TSE, assuming complex interactions among participants' trading activities. 
Our results indicate that early signals of significant price movements can be detected on a daily time scale, offering insights into forecasting structural shifts in financial markets based on time series representing individual participants' behavior.
This study presents a novel perspective by identifying early signals of market instability from individual trading activity, rather than focusing on security price movements, as the conventional analysis subject.
The proposed approach, grounded in participant-level time-series analysis, has potential applications for assessing market conditions and supporting risk management.
Future efforts should refine the choice of time scale for DNM-based forecasting and construct more robust DNM indicators by aggregating various trading time series. 
These developments may evolve the framework into a practical early-warning tool for market instability, with support from collaboration between financial authorities and the academic community for clarifying mechanisms of such market instability.

%%%%%%%%%%%%%%%%%%%%%%%%%%%%%%%%%%%%%%%%%%%%%%%%%%%%%%%%%
\section*{Methods}\label{sec:methods}
In our analysis, we construct various time series $x_i(t)$ representing the trading activity of each participant, and apply the DNM theory to these time series.

\subsection*{Definition of market participant}
We define a market participant as a set of virtual server IDs (VSIDs), considering the trading desk concept and the similarity in trading activities among VSIDs, as follows.

First, we consider VSIDs belonging to the same trading desk as representing the same participant~\cite{goshima2019trader,sato2023inferring,yamada2023_hft_e}. A trading desk, as proposed by Goshima et al. (2019)~\cite{goshima2019trader}, refers to a set of VSIDs that have submitted order placements sharing the same order ID.

Yamada (2023)~\cite{yamada2023_hft_e} outlined criteria for classifying types of trading desks, building on previous studies~\cite{hosaka2014_e,ohyama2021_detailed_data_e,ohyama2022_jafee_e}.
Based on these criteria, we identify HFT-like desks (referred to as {\it HFT desks}) as those satisfying the following conditions:\\
(1) (execution rate)$\leq 25\%$; \\
(2) (cancellation rate)$\geq 20\%$; \\
(3) (proportion of market orders to all orders)$<1\%$;\\
with conditions (1) and (2) satisfied on at least $40\%$ of working days, and condition (3) on at least $60\%$ of working days.
Among trading desks not classified as HFT, those whose PrincipalAgent flag is marked as `self' in the dataset are identified as {\it broker desks}.
Among the remaining desks, those that are neither HFT nor broker desks, satisfy the following conditions on at least $40\%$ of working days to be classified as retail desks (hereafter referred to as {\it general investor desks}:\\
(4) the ratio of margin trading to total execution value is non-zero;\\
(5) the ratio of agency orders, i.e., orders marked as `on behalf of a customer' in the PrincipalAgent flag, to total execution value exceeds $99\%$;\\
(6) over $99.5\%$ of short-selling orders by execution value are conducted through margin trading;\\
(7) the average lifetime from new order submission to cancellation exceeds $5$ seconds;\\
(8) the execution value of the trading desk accounts for at least $0.1\%$ of the total daily execution value.\\
Consequently, we classified $2,136$ desks from November 2019 to March 2020 into $532$ HFT desks, $373$ broker desks, $146$ general investor desks, and $1,085$ others.

Second, we perform hierarchical clustering of VSIDs based on order placements and execution characteristics.
Ohyama et al. (2022)~\cite{ohyama2022_jafee_e} demonstrated that some VSIDs exhibit nearly identical order volumes for certain types of orders, indicating these VSIDs are operated by the same participant using similar trading strategies.
In light of this finding, we applied hierarchical clustering using Ward's method to group VSIDs that appeared between November 2019 and September 2020, based on similarities in trading behavior, with the number of clusters set to $250$ in accordance with Ohyama et al. (2022).
We constructed a characteristic vector $\mathbf{e}^{d,s}$ for each day $d$ and VSID $s$, where entries represent the volumes of order placements and executions of each of $32$ item types shown in Table in Section S.1 of Supplementary Information, along with order placement volumes for each of $4,164$ securities.
Accordingly, the vector $\mathbf{e}^{d,s}$ is $4,196(=4,164+32)$-dimensional.
The distance between two VSIDs, $s$ and $s'$, is calculated as:
\begin{equation}
\sqrt{\displaystyle\sum_{d=1}^D\sum_{k=1}^K (e^{d,s}_k - e^{d,s'}_k)^2},
\end{equation}
where $D=282$ is the number of days analysed and $\mathbf{e}^{d,s}=\mathbf{0}$ if VSID $s$ was inactive on day $d$.

Finally, we determine each participant based on VSID clustering results, with slight modifications to account for trading desks and their types---namely, HFT, broker, general investor, and others.
The initial VSID clustering, performed without considering trading desk information, aligned closely with the groups defined by trading desks: VSIDs from only $60$ out of $2,033$ trading desks were split across multiple clusters. 
In most cases, VSIDs from the same trading desk were assigned to the same cluster.
To resolve these few inconsistencies, we manually reassigned VSIDs in the affected $60$ trading desks, grouping all VSIDs within the same trading desk into the same cluster.
This resulted in a revised cluster structure $\{C_1,...,C_{198}\}$, where $C_i$ denotes the $i$-th cluster of VSIDs.
Following this adjustment, $182$ out of $198$ clusters were composed exclusively of VSIDs from a single trading desk type.
We define each of these $182$ homogeneous clusters as representing an individual participant.
Additionally, we determine a cluster, referred to as {\it left-HFT}, by collecting VSIDs from HFT desks not included in the homogeneous clusters.
Similarly, we defined the clusters {\it left-broker}, {\it left-general-investor}, and {\it left-others}, each of which is regarded as a distinct participant.
Consequently, we identified a total of $186$ participants: $68$ HFTs, $25$ brokers, $3$ general investors, and $90$ others.

\subsection*{Construction of time series representing trading behavior}
We construct a time series $x_i(t)$ for each participant $i$, separately for each trading session (morning or afternoon) and for each day.
Each session lasts $2.5$ hours, with a time unit of one minute, yielding a time series of length $T:=150$ (minutes).

We categorize time series into three groups to represent each participant's trading behavior. Within each category, we analyse nine distinct time series, defined by the type of order placement or execution: 
(1)  trading volumes per minute (vol1--9, Table~\ref{tab:time_series_vol}); 
(2)  participant's centrality within the co-trading network, reflecting overlaps in traded securities among participants (co1--9, Table~\ref{tab:time_series_co});
(3) trading point process, representing the sequence of times at which individual trading activities occur (pp1--9, Table~\ref{tab:time_series_pp}), as follows.

Type (1) time series captures the total trading volume recorded every minute for each participant.
We analyze nine distinct time series within this category, classified by the order placement or execution, as summarized in Table~\ref{tab:time_series_vol}.
For example, for the time series vol1, $x_i(t)$ denotes the total volume of new buy market orders submitted by participant $i$ during the interval $[t,t+1)$ (minutes).

Type (2) time series captures the importance of each participant in co-trading relationships at each time point. 
Prior studies have demonstrated that strong cross-correlations emerge during periods of market instability, observed in stock prices and index prices~\cite{bonanno2001high,kwapien2012physical,Podobnik2010EPL,junior2012correlation,zheng2012changes}.
These time series characterize the microscopic mechanism underlying such cross-correlations by representing the overlap of traded securities across participants.
For each type of trading behavior, from co1 to co9, shown in Table~\ref{tab:time_series_co}, $y_{i,\sigma}(\tilde{t})=1$ indicates the participant $i$ traded security $\sigma$ during the interval $[\tilde{t}, \tilde{t}+1)$ (seconds), and $0$ otherwise.  
We considered $547$ securities that recorded over $10,000$ observations per day throughout the analyzed period.
The number of overlapping securities traded by participants $i$ and $j$ during $[\tilde{t}, \tilde{t}+1)$ is given by
\begin{equation}
z_{i,j}(\tilde{t}) = \sum_{\sigma}y_{i,\sigma}(\tilde{t})y_{j,\sigma}(\tilde{t}).
\end{equation}
Furthermore, we calculated the centrality of participant $i$ in this co-trading relationship, $\sum_j z_{i,j}(\tilde{t})$, at time $\tilde{t}$, aggregated over one minute as
\begin{equation}
x_i(t) = \sum_{\tilde{t}=1}^{60}\sum_j z_{i,j}(\tilde{t}),
\end{equation}
to construct a time series of type (2).
Additionally, we constructed another set of time series, from co1$'$ to co9$'$, where the co-trading relationship is weighted by trade volume (Table~\ref{tab:time_series_co}).
In these time series, $y_{i,\sigma}(\tilde{t})$ represents the aggregated trade volume during the interval $[\tilde{t}, \tilde{t}+1)$ (seconds).

Type (3) time series characterize the statistical properties of the trading point process for each participant.  
A point process is defined as a sequence of times at which individual events occur~\cite{bacry2015hawkes,hawkes2018hawkes,karsai2018bursty}.  
We construct a conventional time series, from pp1 to pp9 in Table~\ref{tab:time_series_co}, by quantifying the heterogeneity of sub-point processes within non-overlapping one-minute windows.
For instance, in constructing the pp1 time series, we consider the point process $\{t_j^i; j=j_1,...,j_m, t\leq t_{j_1}^i\leq\cdots\leq t_{j_m}^i < t+1\}$, which denotes the times at which new buy market orders are submitted by participant $i$ within the interval $[t, t+1)$.
The heterogeneity of this sub-point process is measured by the burstiness indicator $B$~\cite{karsai2018bursty,kim2016measuring}:
\begin{equation}
B=\frac{\sqrt{m+2}\mathrm{CV}-\sqrt{m}}{\left(\sqrt{m+2}-2\right)\mathrm{CV}+\sqrt{m}},
\label{eq:burst}
\end{equation}
where $\mathrm{CV}$ is the coefficient of variation of the inter-event intervals $\Delta_j^i:=t_j^i-t_{j-1}^i$~\cite{karsai2018bursty,Hirata2012PhysA},
\begin{equation}
\mathrm{CV}=\frac{\sqrt{\mathrm{Var}[\Delta_j^i]}}{\langle \Delta_j^i\rangle}.
\end{equation}
In particular, we employed a modified version of the burstiness indicator that accounts for the finite-size effect associated with the number of events $m$.
The value of $x_i(t)$ is then given by $B$ (Eq.~\ref{eq:burst}).

Finally, we pre-processed these time series for DNM analysis to address the strong heterogeneity in magnitudes across participants and the temporal fluctuations within individual time series.
Due to significant differences in magnitudes of $x_i(t)$ between participants (Fig.~\ref{fig:xit}), each time series was normalized by its mean value, $\tilde{x}_i(t) := x_i(t)/\langle x_i(t)\rangle$, where $\langle x_i(t)\rangle := \sum_{t=0}^T x_i(t)/T$.
The normalized time series exhibited intra-day seasonality, with larger values occurring after the start or before the end of trading sessions, and occasional spikes in response to external news.
To mitigate the effect of this non-stationarity on the DNM analysis, we computed the logarithmic return of $\tilde{x}_i(t)$,
\begin{equation}
\tilde{\tilde{x}}_i(t):=\log\displaystyle\frac{\tilde{x}_i(t)}{\tilde{x}_i(t-1)}.
\end{equation}
For simplicity, we denote $\tilde{\tilde{x}}_i(t)$ by $x_i(t)$ in the main text.

\begin{table}[h!]
  \centering
\caption{Time series capturing the trading volume. In the abbreviation in figures, V, O, E, MO, LO, b, and s represent `volume', `order placement', `execution', `market order', `limit order', `buy', and `sell', respectively.}
  %\begin{tabular}{p{0.8cm}p{1.5cm}p{2.7cm}p{2cm}p{1cm}}
  \begin{tabular}{p{1cm}p{2.5cm}p{3.5cm}p{3.7cm}p{1cm}}
    \hline
Symbol & Abbreviation in figures & Order/Execution & Market/Limit order & Buy/Sell \\ 
\hline
vol1 & V\_O\_MO\_b & New order & Market order & Buy \\
vol2 & V\_O\_MO\_s & New order & Market order & Sell \\
vol3 & V\_O\_LO\_b & New order & Limit order & Buy \\
vol4 & V\_O\_LO\_s & New order & Limit order & Sell \\
vol5 & V\_E\_MO\_b & Execution & Market order & Buy \\
vol6 & V\_E\_MO\_s & Execution & Market order & Sell \\
vol7 & V\_E\_LO\_b & Execution & Limit order & Buy \\
vol8 & V\_E\_LO\_s & Execution & Limit order & Sell \\
vol9 & V\_all\_order & New order, Change, Cancellation & Market order, Limit order & Buy, Sell \\
\hline
\label{tab:time_series_vol}
\end{tabular}
\end{table}

\begin{table}[h!]
  \centering
\caption{Time series focusing on the overlap in traded securities across participants. C and CW represent the overlap in traded securities without and with volume weighting, respectively. The expansions of the symbols used are provided in Table~\ref{tab:time_series_vol}.}
  %\begin{tabular}{p{0.8cm}p{1.5cm}p{2.7cm}p{2cm}p{1cm}p{1cm}} 
  \begin{tabular}{p{1cm}p{2.3cm}p{3.5cm}p{3.7cm}p{1.3cm}p{1cm}} 
    \hline
Symbol & Abbreviation in figures & Order/Execution & Market/Limit order & Weight & Buy/Sell \\ 
\hline
co1 & C\_O\_MO\_b & New order & Market order & 0,1 & Buy \\
co2 & C\_O\_MO\_s & New order & Market order & 0,1 & Sell \\
co3 & C\_O\_LO\_b & New order & Limit order & 0,1 & Buy \\
co4 & C\_O\_LO\_s & New order & Limit order & 0,1 & Sell \\
co5 & C\_E\_MO\_b & Execution & Market order & 0,1 & Buy \\
co6 & C\_E\_MO\_s & Execution & Market order & 0,1 & Sell \\
co7 & C\_E\_LO\_b & Execution & Limit order & 0,1 & Buy \\
co8 & C\_E\_LO\_s & Execution & Limit order & 0,1 & Sell \\
co9 & C\_all\_order & New order, Change, Cancellation & Market order, Limit order & 0,1 & Buy, Sell \\
\hline
co1' & CW\_O\_MO\_b & New order & Market order & volume & Buy \\
co2' & CW\_O\_MO\_s & New order & Market order & volume & Sell \\
co3' & CW\_O\_LO\_b & New order & Limit order & volume & Buy \\
co4' & CW\_O\_LO\_s & New order & Limit order & volume & Sell \\
co5' & CW\_E\_MO\_b & Execution & Market order & volume & Buy \\
co6' & CW\_E\_MO\_s & Execution & Market order & volume & Sell \\
co7' & CW\_E\_LO\_b & Execution & Limit order & volume & Buy \\
co8' & CW\_E\_LO\_s & Execution & Limit order & volume & Sell \\
co9' & CW\_all\_order & New order, Change, Cancellation & Market order, Limit order & volume & Buy, Sell \\
\hline
\label{tab:time_series_co}
\end{tabular}
\end{table}

\begin{table}[h!]
  \centering  
\caption{Time series focusing on the trading point process. P represents point process, and the expansions of the symbols used are provided in Table~\ref{tab:time_series_vol}.}
  %\begin{tabular}{p{0.8cm}p{1.5cm}p{2.7cm}p{2cm}p{1cm}}
  \begin{tabular}{p{1cm}p{2.5cm}p{3.5cm}p{3.7cm}p{1cm}}
    \hline
Symbol & Abbreviation in figures & Order/Execution & Market/Limit order & Buy/Sell \\ 
\hline
pp1 & P\_O\_MO\_b & New order & Market order & Buy \\
pp2 & P\_O\_MO\_s & New order & Market order & Sell \\
pp3 & P\_O\_LO\_b & New order & Limit order & Buy \\
pp4 & P\_O\_LO\_s & New order & Limit order & Sell \\
pp5 & P\_E\_MO\_b & Execution & Market order & Buy \\
pp6 & P\_E\_MO\_s & Execution & Market order & Sell \\
pp7 & P\_E\_LO\_b & Execution & Limit order & Buy \\
pp8 & P\_E\_LO\_s & Execution & Limit order & Sell \\
pp9 & P\_all\_order & New order, Change, Cancellation & Market order, Limit order & Buy, Sell \\
\hline
\label{tab:time_series_pp}
\end{tabular}
\end{table}

\subsection*{DNM indicators and market instability}
Empirical studies adopting the DNM theory have proposed methods to identify the DNM set from a system of confoundedly interacting elements. 
In this study, we consider multivariate time series $\mathbf{X}(t)=(X_1(t), ..., X_N(t))$, assumed to be generated by an underlying dynamical system (Eq.~\ref{eq:dyn_sys}).
In practical DNM analysis, for a subset of elements $U$($\subset \{1,...,N\}$), four types of indicators are considered:
the average fluctuation of time series, $\mathrm{SD}_U$, over the elements in $U$;
the average correlation strength among elements within the set $U$, $\mathrm{PCC}_U$; the average correlation strength between elements in $U$ and those outside of it, $\mathrm{PCC}_{\overline{U}}$; and the composite indicator $D_U$, defined as follows:
\begin{align}
\mathrm{SD}_U &=\displaystyle\frac{\sum_{i\in U}\mathrm{SD}(X_i)}{\# U},\label{eq:SDu}\\
\mathrm{PCC}_U &=\displaystyle\frac{\sum_{i,j\in U, i\neq j}\left|\mathrm{PCC}(X_i,X_j)\right|}{\# U(\# U-1)},\label{eq:PCCu}\\
\mathrm{PCC}_{\overline{U}} &=\displaystyle\frac{\sum_{i\in U, j\in\overline{U}}\left|\mathrm{PCC}(X_i,X_j)\right|}{\# U\#\overline{U}},\label{eq:PCCo}\\
D_U &=\displaystyle\frac{\mathrm{SD}_U \cdot \mathrm{PCC}_U}{\mathrm{PCC}_{\overline{U}}}\label{eq:Du},
\end{align}
where $\mathrm{SD}(X)$ denotes the standard deviation of the random variable $X$, $\mathrm{PCC}(X,Y)$ is the Pearson correlation coefficient between $X$ and $Y$, $\#A$ denotes the cardinality of the set $A$ and $|\cdot|$ represents the absolute value.
If $U$ corresponds to the true DNM set, then as the system approaches the instability, $\mathrm{SD}_U$ should increase markedly, $\mathrm{PCC}_U$ should approach $1$, $\mathrm{PCC}_{\overline{U}}$ should approach $0$, and $D_U$ should take on a large value reflecting collective large fluctuation.
Accordingly, in empirical studies, the DNM set $I$ is often identified as the subset $U$ that maximizes $D_U$ prior to instability, based on historical data.
While $D_I$, which integrates information from all three components, is commonly used in DNM analysis, $\mathrm{SD}_I$ is also frequently emphasized due to its robustness in capturing early warning signals~\cite{masuda2024anticipating,nakagawa2016early,chen2012detecting}.
It is also recommended to remove $\mathrm{PCC}_{\overline{U}}$ for the sake of simplicity~\cite{aihara2022dynamical}.

Among four DNM indicators, we focused on $\mathrm{SD}_I$ due to its robustness against noise and compared its daily values in relation to market instability, as indicated by large price fluctuations.
Since $\mathrm{SD}_I$ is the mean standard deviation of time series across participants in the DNM set, we first calculated the standard deviation of all participants' time series. We selected participants whose standard deviation increased prior to large price fluctuations to determine the elements of the DNM set.

For robust comparison of the standard deviation of the time series and market instability, we smoothed the daily standard deviation and evaluated it relative to the previous week.
We calculated the standard deviation of each time series.
The standard deviation of the morning session on day $d$ for participant $i$ is given by
\begin{equation}
\mathrm{SD}_{i,\mathrm{am}}(d) := \sqrt{\frac{1}{T}\sum_{t=0}^T\Bigl(x_i(t)-\langle x_i(t)\rangle\Bigr)^2}.
\end{equation}
The distributions of these values are shown in Section S.2 in Supplementary Information.
Similarly, the standard deviation for the afternoon session is denoted by $\mathrm{SD}_{i,\mathrm{pm}}(d)$.
The values of $\mathrm{SD}_{i,\mathrm{am}}(d)$ and $\mathrm{SD}_{i,\mathrm{pm}}(d)$ exhibited noisy fluctuations that obscured the underlying trend.
To reduce noise, we employed a moving average over five consecutive working days (approximately one week), averaging the morning and afternoon values from each day to yield ten observations in total:
\begin{equation}
\overline{\mathrm{SD}_i(d)} := \frac{1}{10}\sum_{\tau=0}^4\Bigl(\mathrm{SD}_{i,\mathrm{am}}(d+\tau) + \mathrm{SD}_{i,\mathrm{pm}}(d+\tau)\Bigr).
\end{equation}
We then defined the logarithm of the relative standard deviation (Log-RSD) $R_i(d)$ as
\begin{equation}
R_i(d) := \log\displaystyle\frac{\overline{\mathrm{SD}_i(d)}}{\overline{\mathrm{SD}_i(d-5)}}.
\end{equation}
This ratio normalizes differences in magnitude across participants and highlights increases in $\overline{\mathrm{SD}_i(d)}$~\cite{song2024early}.

In this study, daily price movements are measured using the Tokyo Stock Price Index (TOPIX).
The analysed dataset is the TOPIX Tick Data provided by the JPX Data Cloud (\url{https://db-ec.jpx.co.jp/}), which records the index value, $\tilde{V}_d(t)$, every second.
We defined the volatility, $V(d)$, on day $d$ as the standard deviation of the logarithmic returns of TOPIX values as $V(d):=\text{Std}[\tilde{\tilde{V}}_d(t)]$, where $\tilde{\tilde{V}}_d(t) := \log\tilde{V}_d(t) - \log\tilde{V}_d(t-1)$.
We focus on $V(d)/V(d-1)$, referred to as Relative Volatility, which captures the onset of instability rather than its persistence.
To examine the performance of the early-warning signal in detecting significant episodes of large price movements, we then identified the five days on which Relative Volatility exhibited the highest values during the observation period.
These {\it turmoil days} are 8 January, 25 February, 9 March, 13 March, and 28 August; consecutive days were excluded from the selection.
On these days, other volatility indicators also exhibited elevated values compared with the others, such as $\mathrm{E}[|\tilde{\tilde{V}}_d(t)|]$, the difference between the closing and opening values of the TOPIX index on consecutive days, and the difference between consecutive closing values.

\subsection*{Co-trading network}
We defined principal participants as those appearing repeatedly in the DNM set across multiple time-series types and focal turmoil days, examining their position in the co-trading network, where nodes represent participants and links indicate a strong co-trading relationship.
For each pair $(i,~j)$ of participants, $z_{i,j}(d)$ denotes the number of shared securities ordered on day $d$, and $w_{i,j}(=\sum_d z_{i,j}(d))$ gives the cumulative weight in this month.
Backbone extraction was applied to the weighted network, retaining links satisfying $\left(1-w_{i,j}/s_i\right)^{k_i-1}<0.01$ or $\left(1-w_{i,j}/s_j\right)^{k_j-1}<0.01$, where $k_i$ and $s_i$ denote the degree and strength of node $i$ ($s_i=\sum_j w_{i,j}$)~\cite{menczer2020first}.

\section*{Abbreviations}
\noindent {\bf DNM:} Dynamical network marker\\
\noindent {\bf VSID:} Virtual server ID\\
\noindent {\bf HFT:} High frequency trading\\
\noindent {\bf Log-RSD:} Log-Relative Standard Deviation

%%%%%%%%%%%%%%%%%%%%%%%%%%%%
%%%%%  References %%%%%%%%%%
%%%%%%%%%%%%%%%%%%%%%%%%%%%%
%\bibliographystyle{naturemag} 
%\bibliography{reference} 

\noindent{\bf Supplementary information}\\
Supplementary Information provides detailed results on the clustering of server IDs and a complete set of results corresponding to the analyses presented in the main text, covering all time series and all months, whereas the main text presents only a subset for illustrative purposes.

%\section*{Declarations}
\noindent{\bf Availability of data and materials}\\
The data that support the findings of this study are provided by Japan Exchange Group and are not publicly available due to confidentiality restrictions. The data were used under license for the current study. Data may be made available from the corresponding author upon reasonable request, subject to approval by Japan Exchange Group and Financial Services Agency, Japan.

\noindent{\bf Competing interests}\\
The authors declare no competing interests.

\noindent{\bf Funding}\\
This work was supported by Project Fund for Center for Social Complex Systems, Institute of Industrial Science, the University of Tokyo, and partially supported by JSPS KAKENHI Grant Numbers JP22H01719 and JP25K15349, JST Moonshot R\&D Grant Number JPMJMS2021, a project, JPNP14004, commissioned by the New Energy and Industrial Technology Development Organization (NEDO), and Institute of AI and Beyond of UTokyo. 

\noindent{\bf Author contributions}\\
M.I.I., Y.H., T.O., T.W., and K.A. conceived and designed the study. M.I.I. and H.H. performed the investigation and analyzed the data. M.I.I., H.H., Y.H., and T.O. developed the methodology and validated the results. Y.H. administered the project. M.I.I. drafted the manuscript. All authors reviewed and approved the final manuscript.

\noindent{\bf Acknowledgements}\\
The authors would like to thank  Prof. Naoyuki Yoshino and Dr. Atsuyuki Ohyama, as well as members of the Financial Service Agency, Japan, for their valuable comments and helpful discussions.

\clearpage
%%%%%%%%%%%%%%%%%%%%%%%%%%%%
%%%%%%%%  Figures %%%%%%%%%%
%%%%%%%%%%%%%%%%%%%%%%%%%%%%

\begin{figure}[h]%[tb]
\centering
\includegraphics[width=0.83\textwidth]{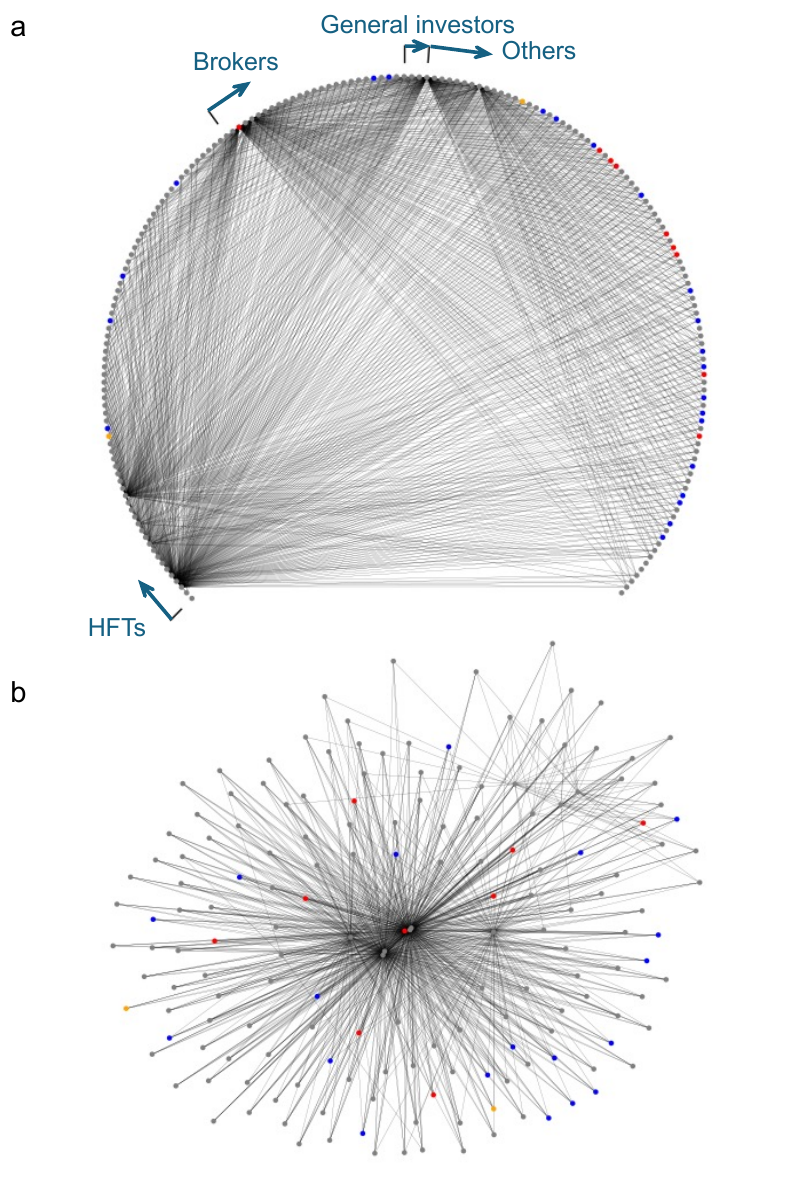}%[width=12.5cm]{fig6}%Fig/fig_principal_participants.pdf}
\caption{{\bf Network representation of co-trading relationships among participants.} Each node represents a participant, and each link indicates a strong co-trading relationship. The set of principal participants identified by considering all time-series types (DNM1) and those identified using only types associated with trading point processes (DNM2) are highlighted in the network. Principal participants belonging only to DNM1 and DNM2 are shown in blue and red, respectively, while those belonging to both are shown in orange. In panel (a), participants are arranged clockwise in the same order as on the horizontal axis in Fig.~\ref{fig:marketparticipants_all}. In panel (b), nodes are arranged such that tightly connected nodes are placed close to each other.}
\label{fig:stock_sharing_network}
\end{figure}

\clearpage

%\includepdf[pages={-},clearpage]{SI_submit.pdf}
\includepdf[pages=-]{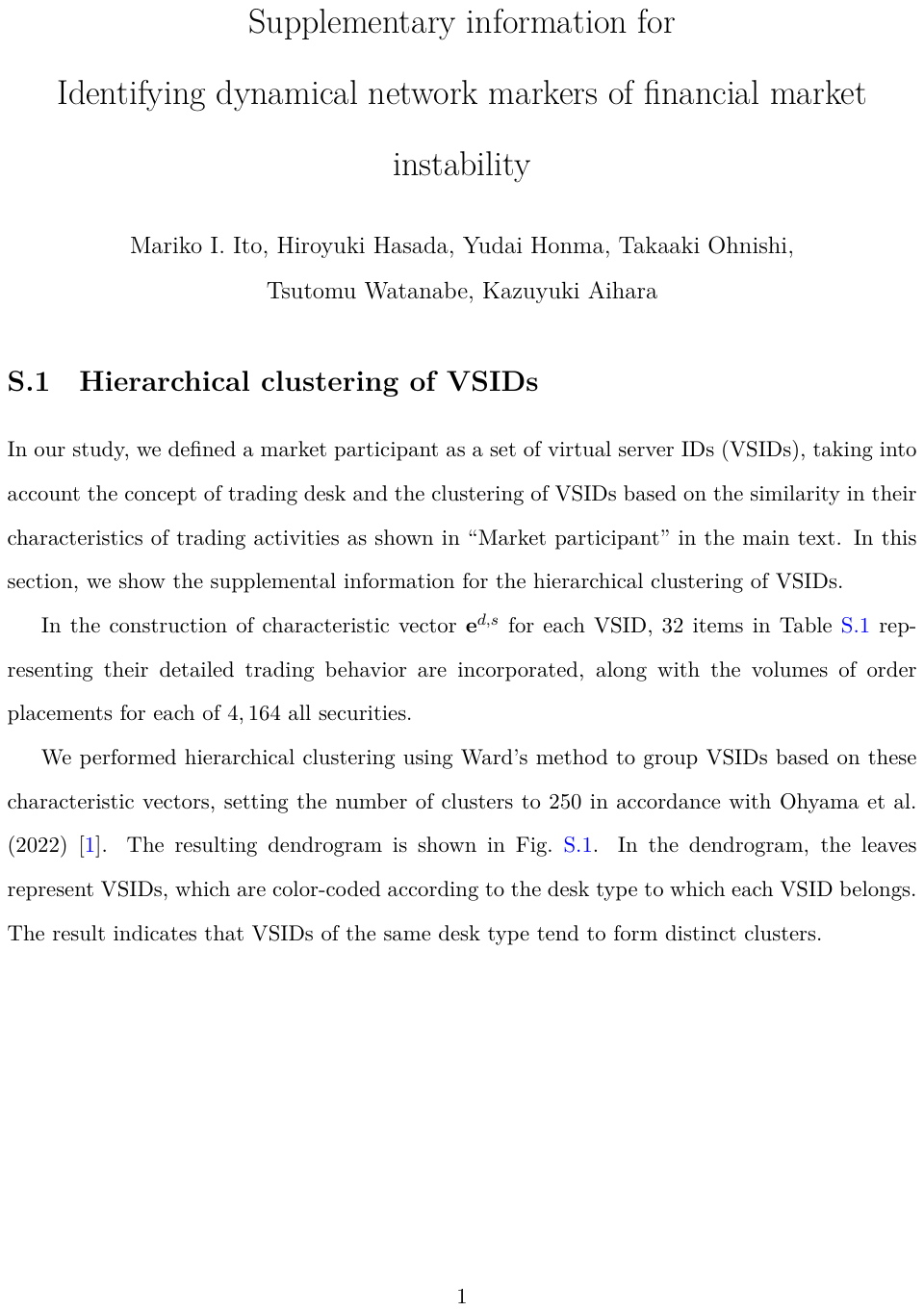}

\end{document}